\documentclass[aps,prl,twocolumn,showpacs,10pt,superscriptaddress,floatfix,longbibliography]{revtex4-1}

\usepackage[usenames, dvipsnames]{color}
\usepackage[normalem]{ulem}
\usepackage{amsmath}
\usepackage{float}
\usepackage{color}
\usepackage{bbold}
\usepackage{amssymb}
\usepackage{amsfonts}
\usepackage{bm}
\usepackage[edges]{forest}

\usepackage{rotating}

\usepackage[normalem]{ulem}
\usepackage{soul}
\usepackage{amssymb}
\usepackage{graphicx}

\usepackage[colorlinks, breaklinks, 
linkcolor=OrangeRed,
citecolor=RoyalBlue,
urlcolor=RoyalBlue]{hyperref}          
\usepackage{multirow}
\usepackage[all]{hypcap} 
\usepackage{xspace}
\usepackage{amssymb}
\usepackage{appendix}
\usepackage{pifont}

\usepackage{etoolbox} 
\usepackage{lipsum} 
\usepackage[capitalize]{cleveref}

\makeatletter
\appto{\appendix}{%
  \@ifstar{\def\theequation@prefix{A.}}%
          {}%
}
\makeatother

\begin{document}

\title{Topological Optical Pseudospin Injection Beyond Weyl Semimetals}

\author{Suvendu Ghosh\textsuperscript{\textcolor{blue}{\textdaggerdbl}}}
\email{suvenduphys@gmail.com}
\thanks{\textcolor{blue}{\textdaggerdbl}~Both authors SG and CZ contributed equally.}
\affiliation{Department of Physics, Indian Institute of Technology Kharagpur, Kharagpur 721302, India}

\author{Chuanchang Zeng\textsuperscript{\textcolor{blue}{\textdaggerdbl}}}
\affiliation{Beijing Academy of Quantum Information Sciences, Beijing, 100193, China} 

\author{A. Taraphder}
\affiliation{Department of Physics, Indian Institute of Technology Kharagpur, Kharagpur 721302, India}

\author{Jian-Xin Zhu}
\affiliation{Theoretical Division, Los Alamos National Laboratory, Los Alamos, New Mexico 87545, USA}
\affiliation{Center for Integrated Nanotechnologies, Los Alamos National Laboratory, Los Alamos, New Mexico 87545, USA}

\author{Snehasish Nandy}
\email{snehasish@phy.nits.ac.in}
\affiliation{Department of Physics, National Institute of Technology Silchar, Assam 788010, India}
\affiliation{Theoretical Division, Los Alamos National Laboratory, Los Alamos, New Mexico 87545, USA}

\begin{abstract}
Photoinduced effects are now reckoned to be important tools to reveal a rich gamut of entrancing physics in topological materials, which are normally inaccessible to conventional probes. Here we investigate one of these intriguing effects, namely, {\it optical pseudospin injection} (OPI) beyond ordinary Weyl semimetals (WSMs), specifically in multi-WSMs (mWSMs) and higher-order WSMs. Remarkably, we demonstrate that OPI in mWSMs is independent of the frequency of the light and linearly proportional to the quantized topological charge as a consequence of the inherent band linearity in their dispersions. Interestingly, while the response does not depend on the tilting of a type-I node, it is a decreasing function of the same in type-II mWSMs. We also reveal that the frequency-independence can be destroyed either by going beyond a certain cutoff frequency under lattice regularization or by going to a higher-order Weyl phase. The predicted signatures of OPI beyond the ordinary WSM could be experimentally exploited, leading to effective access as well as distinguishing between different nontrivial Weyl topologies.

\end{abstract}

\maketitle

{\color{blue}{\em Introduction}}--- Three-dimensional Weyl semimetals (WSMs) constitute a fascinating class of topological materials, where the valence and conduction bands touch at isolated points, called Weyl nodes (WNs), in the Brillouin zone~\cite{Mele_2018, Murakami_2007, Leon_2011}. The low-energy excitations around a WN are effectively described by the Hamiltonian $H(\mathbf{k})=s\hbar\sum_i v_i \sigma_i k^i$, where $\hbar$ is the reduced Planck constant, $v_i$'s are band velocities, and $\sigma_i$'s are Pauli matrices representing pseudospin structure intrinsic to WNs~\cite{Mele_2018}. The winding number or the topological charge ($\pm n$), fundamental to the topological characteristics of WNs, is then defined by the winding of the pseudospin vector around a node in momentum space~\cite{Zhao16, Yang16, Elbistan17, Samokhin19}, and is quantized to integer values $\pm 1$ for WSMs. The sign ($\pm$) represents the chirality index ($s$) of a node, the conservation of which requires WNs to always arise in pairs and act as sources or sinks of Berry curvature in momentum space~\cite{Ninomiya_1981, Ninomiya_1983, Mele_2018, Niu_2010}. 

Interestingly, proposals have emerged to realize topological semimetals hosting WNs with higher topological charges ($n>1$), specifically referred to as multi-Weyl semimetals (mWSMs)~\cite{Zhong_2011, Fang_2012, Huang_2016, Yang_2014}. Unlike typical WSMs with isotropically linear energy dispersion in momenta, mWSMs with $1<n \le 3$, constrained by point-group symmetries such as $C_4$ and $C_6$ rotational symmetries~\cite{Fang_2012, Yang_2014}, display naturally anisotropic dispersion. More recently, based on symmetry and momentum-resolved topological invariants, a novel class of topological phases, dubbed as {\it higher-order topological phases}, has been proposed to exist in condensed matter systems~\cite{Schindler18, Roy_2019, Wieder20, Wu20, Sayed_2020, Hua_2020, Agarwala_2020, Noguchi21, Luo21, Wei21, Xie21, Yu22}. An $p^{th}$ order topological phase is characterized by the nonzero higher-order topological invariant - $2^p$ moment (e.g., quadrupole for $p = 2$) in the bulk and hosts $(d-p)$ dimensional topologically protected states on its edges or corners where $d$ is the dimension of the system~\cite{Schindler18, Roy_2019}. One of the latest additions to this class is {\it higher-order Weyl semimetal }(HOWSM)~\cite{Sayed_2020,Hua_2020,Luo21,Wei21}. A second-order WSM features at least a pair of second-order Weyl nodes of opposite quantized charge within its bulk, resulting in gapless hinge states with {\it fractional} charge. Notably, a hybrid-order WSM may also arise, with coexisting first- and second-order nodes in the bulk, thereby giving rise to both hinge states and Fermi arcs on the surface~\cite{Sayed_2020, Hua_2020}.

Owing to their unusual band topology, WSMs and mWSMs showcase a variety of fascinating optical phenomena~\cite{Mele_2018, Nagaosa20, Orenstein21, Guo23, Nandy_2022, Dey_2020, Banashree_2021, Lee_2017, Zhang_2018, Min-2016, Tewari_2015, Cao24}. It has been recently proposed that the helical spin-momentum locking around WNs can give rise to a frequency-independent circular optical spin injection in ordinary WSMs~\cite{Gao24} and thereby makes ordinary WSMs promising for spintronics~\cite{Sarma_2004}. However, for the sake of generalization, we extend the concept to any spin-like degrees of freedom, i.e., pseudospins in these systems, which might lead to the generation of pseudospin-polarized electrons~\cite{Ye17, Bertrand19, Parameswaran14, Murotani24} by circularly polarized light illumination, a phenomenon dubbed as circular {\it optical pseudospin injection (OPI)}. Therefore, some pressing questions immediately come to the fore: what implications would the circular OPI hold for WNs with higher topological charges and in higher-order Weyl phases? Particularly intriguing is the possibility of probing different $n$ in such systems and discerning them experimentally — an area of profound fundamental interest. It would also be tempting to know whether the frequency-independence of the response persists beyond an ordinary WSM, specifically in mWSMs and HOWSMs.

In this Letter, in light of the above questions we investigate circular OPI in mWSMs and HOWSMs. Interestingly, we find that the trace of the OPI coefficient in mWSMs is {\it independent of the frequency} of the circularly polarized light and {\it linearly proportional} to the quantized topological charge associated with a node. We attribute these striking features of OPI to the inherent {\it band linearity} along only one direction in the dispersion of mWSMs. This stands in sharp contrast to the previous claim of a frequency-independent response as a product of frequency-independent helicity and chirality~\cite{Gao24}. We show that, unlike a type-I node, the circular OPI is a decreasing function of the tilting parameter in type-II mWSMs. Interestingly, we reveal that the frequency-independence can be disrupted in two ways: (i) by transitioning to a higher frequency range due to the appearance of nonlinearity in the band dispersion, and (ii) by going beyond the first-order topological phase (even within the low-frequency range). We believe the predicted signatures of OPI in this work could serve as a potential diagnostic tool to identify different Weyl phases in experiments and would render a new avenue for contemporary technological applications.

{\color{blue}{\em Circular optical pseudospin injection and its general symmetry requirements}}--- The circular OPI, a second-order optical phenomenon, refers to the generation of pseudospin-polarized electrons in the conduction band due to interband transitions in response to a circularly polarized light (CPL) illumination~\cite{Meier_1984, Sarma_2004, Najmaie_2003, Sipe_2007, Sipe_2014}. These transitions are determined by the optical selection rules and the resulting rate of pseudospin generation 
can be expressed as~\cite{Choi_2017}
\begin{eqnarray}
\frac{dm_\alpha}{dt}=\gamma_{\alpha \beta}(\omega)[i\mathbf{E}(\omega) \times \mathbf{E^{*}}(\omega)]_{\beta}\;.
\end{eqnarray}
Here $\mathbf{m}$ is the pseudospin magnetization in the system, and $\mathbf{E}(\omega)$ is the Fourier component of the electric field at an angular frequency $\omega$. The indices $\alpha$ and $\beta$ stand for directional components of the pseudospin magnetization and the electric field respectively. Clearly, the rate of pseudospin generation is proportional to the light intensity. The quantity $\gamma_{\alpha \beta}(\omega)$ is the optical pseudospin injection tensor. Interestingly, its diagonal components ($\gamma_{\alpha \alpha}$) do not need to break any mirror, spatial inversion, or time-reversal symmetry because both $\mathbf{m}$ and $\mathbf{E}(\omega) \times \mathbf{E^{*}}(\omega)$ (since $\mathbf{E}$ is a polar vector) transform as a pseudovector. Therefore, the trace of $\gamma_{\alpha \beta}$ transforms as a scalar under point group operations.

To obtain the general expression of the diagonal part of the injection coefficient, we first consider the light-electron interaction Hamiltonian as $H_{le}=-e\, \mathbf{v}\cdot\mathbf{A}$, where  $e$ is the charge of the electron, $\mathbf{A}$ is the vector potential of the light and $\mathbf{v}$ is the velocity operator. We neglect Zeeman coupling with the light field because the Zeeman coupling is found to be weaker than the orbital contribution for typical Dirac and Weyl systems. Considering the second-order response theory within the density matrix formalism on the total Hamiltonian $\hat{H}=\hat{H}_0+\hat{H}_{le}$ with $\hat{H}_0$ being the unperturbed Hamiltonian, the general expression of $\gamma_{\alpha \alpha}$ can be written as ($e=\hbar=1$ for simplicity)~\cite{Gao20, Gao24}
\begin{eqnarray}
\gamma_{\alpha \alpha}=-\frac{\tau g}{8\pi^2} \sum_{l,m} \int d^3k (\Omega_\alpha)_{ml} (\Delta s_\alpha)_{ml}\Delta f_{lm} \delta(\omega_{lm}-\omega), \nonumber \\
\label{res_fun_F}
\end{eqnarray}
where $\tau$ is the relaxation time, $g$ is the effective g-factor, and the indices $m$ and $l$ represent the band indices. Here, $\omega_{lm}=\epsilon_{l}-\epsilon_{m}$, and $\Delta f_{lm}=f_{l}-f_{m}$  with $f_l$ be the Fermi-Dirac distribution function for band $l$. The quantity $(\Omega_\alpha)_{ml}=\epsilon_{\alpha \beta \delta} \rm Im[\langle u_m|i\partial_{k_\beta}|u_l\rangle \langle u_l|i\partial_{k_\delta}|u_m\rangle]$ is the band-resolved Berry curvature with $|u_l\rangle$ be the periodic part of the Bloch wavefunction. The $\mathbf{k}$-resolved pseudospin difference between two bands, $(\Delta s_\alpha)_{ml}$, is expressed  as $(\Delta s_\alpha)_{ml}=\langle u_m|s_\alpha|u_m\rangle -\langle u_l|s_\alpha|u_l \rangle$. 

We would like to point out that the OPI, in general, can have different origins. In the current work, we focus on the contribution induced by the Abelian Berry curvature, as given in Eq.~(\ref{res_fun_F}), as the topology of the WSMs and mWSMs is directly connected to monopole and anti-monopole of the Abelian Berry curvature. However, since the Abelian Berry curvature distribution vanishes in the presence of both inversion and time-reversal symmetries (e.g., centrosymmetric nonmagnetic systems), the diagonal component of the OPI tensor will vanish. In that case, optical injection of spin or pseudospin might occur through different mechanisms such as proximity coupling~\cite{Avsar2017}, application of external magnetic field~\cite{Tiwary10, Inglot14}, optical injection of chirality~\cite{Murotani24}. It is important to note that the OPI differs from the inverse Faraday effect (IFE) and circular photogalvanic injection (CPGI) current~\cite{Oppeneer_2014, Tokman_2020, deJuan_2017}. The IFE describes a light-induced static magnetization ($\mathbf{m}$) linearly proportional to the light intensity in contrast to a dynamical magnetization ($\propto \frac{d\mathbf{m}}{dt}$) in OPI.


{\color{blue}{\em Optical pseudospin injection in mWSMs}}--- To study the OPI in mWSM, we consider a generic model of a multi-Weyl node of chirality $s$ with broken inversion and time-reversal symmetries. The corresponding low-energy Hamiltonian can be written as~\cite{Li16,Mukherjee18,Ghosh24}
\begin{eqnarray}
&&H_{n}^{s} \left( \mathbf{k} \right) =\nonumber \\
&&s\left[\alpha_{n} k^n_{\bot} \left[ \cos \left( n \phi_{k} \right) 
\sigma_{x} +\sin \left( n \phi_{k} \right) \sigma_{y} \right] + v_z (k_z-sQ) \sigma_{z}\right]\nonumber \\
&& + C_s v_z(k_z-sQ)-sQ_0,
\label{eq_multi1}
\end{eqnarray}
where $n$ is the topological charge, 
$k_{\bot}=\sqrt{k_x^2+k_y^2}$, and $\phi_k={\rm arctan}(k_y/k_x)$. Here, $\sigma_x,\, \sigma_y$, and $\sigma_z$ are Pauli matrices acting on the pseudospin basis, and $s\,(=\pm)$ is the chirality index of a multi-Weyl node. Here, $v_z$ and $C_s$ denote the velocity and tilt parameter along the $z$-direction respectively. $Q$ ($Q_0$) represents the momentum space (energy axis) coordinate of the node and breaks the time-reversal (inversion) symmetry. Also, $\alpha_n=\frac{v_{\perp}}{k_0^{n-1}}$, where $v_{\perp}$ is the effective velocity of the quasiparticles in the plane perpendicular to the $z$ axis and $k_0$ represents a material-dependent parameter having the dimension of momentum. 

The resulting energy dispersion of a multi-Weyl node associated with chirality $s$ is given by $\epsilon_{\mathbf{k},s}^{\pm} =C_s v_z (k_z-sQ)-sQ_0 \pm \sqrt{ \alpha^2_{n} k^{2 n}_{\bot} + v_z^2 (k_z-sQ)^2}$,  where $\pm$ represents conduction and valence bands respectively. The dispersion around a single Weyl node ($n=1$) is isotropic in all momentum directions for $v_z=v_{\perp}=v$. On the other hand, for $n=2$ $(3)$, we find that the dispersion around a double (triple) Weyl node becomes quadratic (cubic) along both $k_x$ and $k_y$ directions whereas varies linearly with $k_z$. Based on the tilt parameter strength $C_s$, a Weyl node can be classified as type-I or type-II. Specifically, type-I and type-II (over-tilted) multi-Weyl nodes are characterized by a point-like Fermi surface ($|C_s|<1$) and finite Fermi pocket ($|C_s|>1$) at the node~\cite{Soluyanov15, Jin22}, respectively. Interestingly, in contrast to a type-I node, a type-II node breaks the Lorentz symmetry in mWSM~\cite{Xu17}. 

Different Berry curvature components of a multi-Weyl node are given by $\mathbf{\Omega}_{\mathbf{k},s}^{\pm} =\pm \frac{s}{2} \xi(\mathbf{k})\: \{ k_x, k_y, n (k_z-sQ) \}$, where $\xi(\mathbf{k})=\frac{n v_z \alpha_n^2 k^{2n-2}_{\bot} }{\beta_{\mathbf{k},s}^{3}}$ and $\beta_{\mathbf{k},s}=\sqrt{ \alpha^2_{n} k^{2 n}_{\bot} + v_z^2 (k_z-sQ)^2}$. Similar to energy dispersion, the Berry curvature is isotropic in all momentum directions for an ordinary Weyl case whereas it becomes anisotropic for WSMs with $n>1$, i.e., for double WSM ($n=2$) and triple WSM ($n=3$). 

Having calculated the Berry curvature, we now investigate the OPI in mWSM. It is important to note that we ignore internode coupling phenomena and apply the approach of independent node approximation~\cite{Cote23}. Therefore, the total contribution becomes the sum of the individual contributions from all the nodes. Following Eq.~(\ref{res_fun_F}), the trace of the OPI tensor for a single node with chirality $s$ takes the form
\begin{equation}
\rm{Tr}\gamma_{\alpha\beta}^{s}=-\frac{\tau g}{8n\pi^2} \int d^3k \frac{(\mathbf{\Omega}_{\mathbf{k},s}^{+} \cdot \mathbf{v}_{\mathbf{k},s}^{+}) (\mathbf{\tilde{k}}_s\cdot \mathbf{\Delta s}_{\mathbf{k},s})}{\epsilon_{\mathbf{k},s}^{+}+sQ_0} \delta(\omega_{cv}-\omega)
\label{res_fun_mWSM}
\end{equation}
where $\mathbf{\tilde{k}}_s=k_x \, \mathbf{\hat{x}} +k_y \, \mathbf{\hat{y}} +n(k_z-sQ) \, \mathbf{\hat{z}}$. Here $\omega_{cv}=\epsilon_{\mathbf{k},s}^{+}-\epsilon_{\mathbf{k},s}^{-}$ and $\mathbf{\Delta s}_{\mathbf{k},s}=\frac{s}{\beta_{\mathbf{k},s}}\left[\alpha_nk^n_{\perp}\{\cos{n\phi_k} \, \mathbf{\hat{x}}+\sin{n\phi_k} \, \mathbf{\hat{y}}\}+v_z(k_z-sQ) \, \mathbf{\hat{z}}\right]$ are the energy and pseudospin difference between bands respectively. Since both the Berry curvature ($\mathbf{\Omega}_{\mathbf{k},s}$) and the pseudospin difference ($\mathbf{\Delta s}_{\mathbf{k},s}$) are proportional to the chirality index, the response coefficient $Tr\gamma_{\alpha\beta}^{s}$ remains chirality-independent and, consequently, pair-wise contributions do not vanish. 

The trace of the OPI tensor for an inversion and time-reversal symmetry broken type-I mWSM given in Eq.~(\ref{eq_multi1}) is obtained as
\begin{eqnarray}
\rm{Tr}\gamma_{\alpha\beta}\, &&=-n \gamma_0 \left(\frac{2}{v_{\perp}}\delta_{n,1}+\frac{1}{v_z}\right), 
\label{untilt_freq0}
\end{eqnarray} 
where $\gamma_0 = \tau g/12\pi$ and $\delta_{n,1}$ is the Kronecker delta. Equation~(\ref{untilt_freq0}) is one of our main results and has some striking features. {\it First,}  ${\rm Tr}\gamma_{\alpha\beta}$ is {\it independent} of applied light frequency $\omega$. This could have immense implications in pseudospintronic and spintronic applications~\cite{Sarma_2004, Pesin12}, as a frequency-independent OPI implies no requirement of fine-tuning of light frequency close to resonance to generate pseudospin magnetization. {\it Second,} the circular OPI in mWSMs is remarkably tied with the topology of a node. It can be seen from Eq.~(\ref{untilt_freq0}) that ${\rm Tr}\gamma_{\alpha\beta}$ is {\it linearly proportional} to the quantized topological charge $n$ associated with the node. Consequently, the circular OPI response could directly probe the topology of the mWSMs and, thus, serve as an important probe to distinguish an ordinary WSM from an mWSM in experiments. Notably, the OPI response is increasing with $n$ in mWSMs. {\it Third,} the OPI response is independent of the tilting of the node (particle-hole asymmetry), the time-reversal breaking parameter $Q$, and the inversion breaking parameter $Q_0$, ensuring our results to be valid for a large class of type-I WSM phases.

\begin{center}
\begin{figure}
\includegraphics[width=0.47\textwidth]{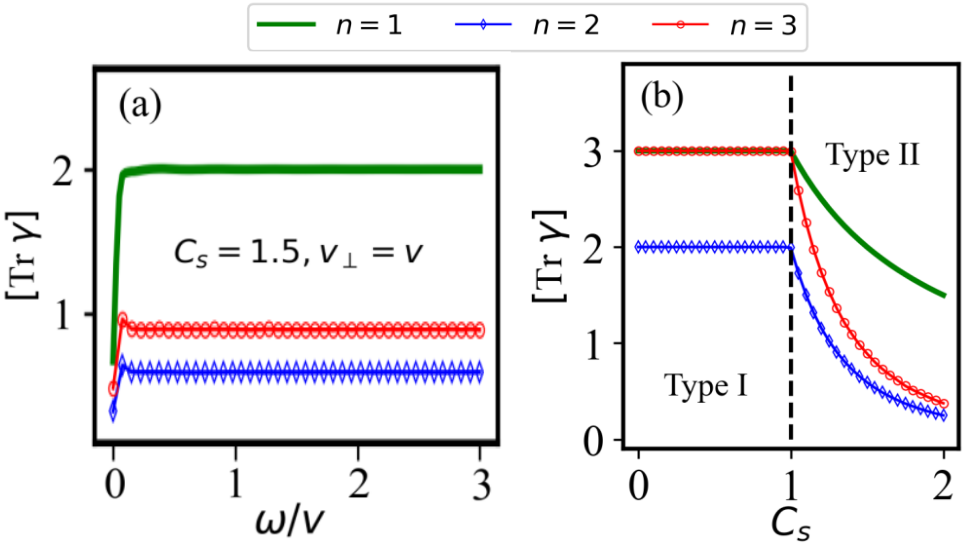}
\caption{The trace of the OPI response based on the low-energy mWSM model. Here, [Tr $\gamma$] is $\rm{Tr}\gamma_{\alpha\beta}$ normalized by $(-\gamma_0/v_z)$. Panel (a): [Tr $\gamma$] for type-II ($C_s=1.5$) mWSMs as a function of light frequency. The $\omega$-independence and the quantization of the response to some values, for every $n$, can be seen. Panel (b): [Tr $\gamma$] as a function of tilt parameter ($C_s$) for a fixed $\omega = 2 v$. The tilt-independence of the response can be seen for type-I nodes. Conversely, it decreases with increasing tilt strength in the type-II regime. The other parameters used here are: $v_{\perp} = v_z = v = 1$, $Q = Q_0 = 0$, and $k_0 = 0.8$. 
}
\label{fig2_analytical}
\end{figure}
\end{center}

To delve into it further, we analytically investigate ${\rm Tr}\gamma_{\alpha\beta}$ component-wise and find distinct origins of OPI response in mWSMs compared to an ordinary WSM. Interestingly, for ordinary WSMs ($n=1$), all three components of $\gamma_{\alpha\alpha}$ ($\alpha=x,\,y$ and $z$) contribute non-vanishingly to ${\rm Tr}\gamma_{\alpha\beta}$ and become equal in an isotropic case ($v_z=v_{\perp}=v$). On the other hand, due to an azimuthal ($\phi_{k}$) symmetry in dispersions of mWSMs around $k_z$-axis, $\gamma_{xx}$ and $\gamma_{yy}$ strikingly vanish for multi-Weyl nodes ($n>1$) and only non-zero contribution comes from $\gamma_{zz}$. This directional dependence is also evident from Eq.~(\ref{untilt_freq0}): ${\rm Tr}\gamma_{\alpha\beta}$ for mWSM depends only upon $v_z$ whereas, for ordinary WSM, it depends upon both $v_{\perp}$ and $v_z$. 

The strong directional dependence of the OPI response could shed light on the possible origin of its frequency-independence and linear proportionality with $n$. We note that, unlike the isotropic distribution of the underlying Berry curvature in the momentum space for an ordinary WSM, the naturally anisotropic band dispersion for any $n>1$ makes the underlying Berry curvature of mWSMs anisotropic. The topological charge ($n$) then becomes associated only with the longitudinal ($k_z$) component of Berry curvature, along which the band dispersion is linear. Therefore, we argue that the frequency-independence and linear proportionality with $n$ of OPI response in mWSM appears due to the band linearity along $k_z$-direction. Interestingly, the OPI coefficient is affected by the anisotropic velocity ratio ($v_{\perp}/v_z$) only in ordinary WSM as seen from Eq.~(\ref{untilt_freq0}). Therefore, by tuning  $v_{\perp}/v_z$ in ordinary WSM, one can achieve the OPI coefficient to be equal with triple-WSM ($n=3$) when $v_z=v_{\perp}=v$ and $\tau^{\rm WSM}=\tau^{\rm mWSM}$. The above results are also verified by the numerical calculation of the low-energy model as shown in Fig.~\ref{fig2_analytical}.

Going beyond the type-I regime, we have investigated the OPI for the type-II regime by introducing an energy cutoff $\epsilon_c\sim\hbar v\Lambda_c$, where $\Lambda_c$ is the corresponding momentum cutoff~\cite{Roy16, Ghosh19}. We find that the $\omega$-independence and linear proportionality with $n$ remain intact also in type-II mWSMs, as can be seen from Fig.~\ref{fig2_analytical}(a). However, the response no longer remains independent of the tilting effect. It can be seen from Fig.~\ref{fig2_analytical}(b) that the magnitude of the response in these cases rapidly dies down as the tilting effect increases. 

\begin{center}
\begin{figure}
\includegraphics[width=0.47\textwidth]{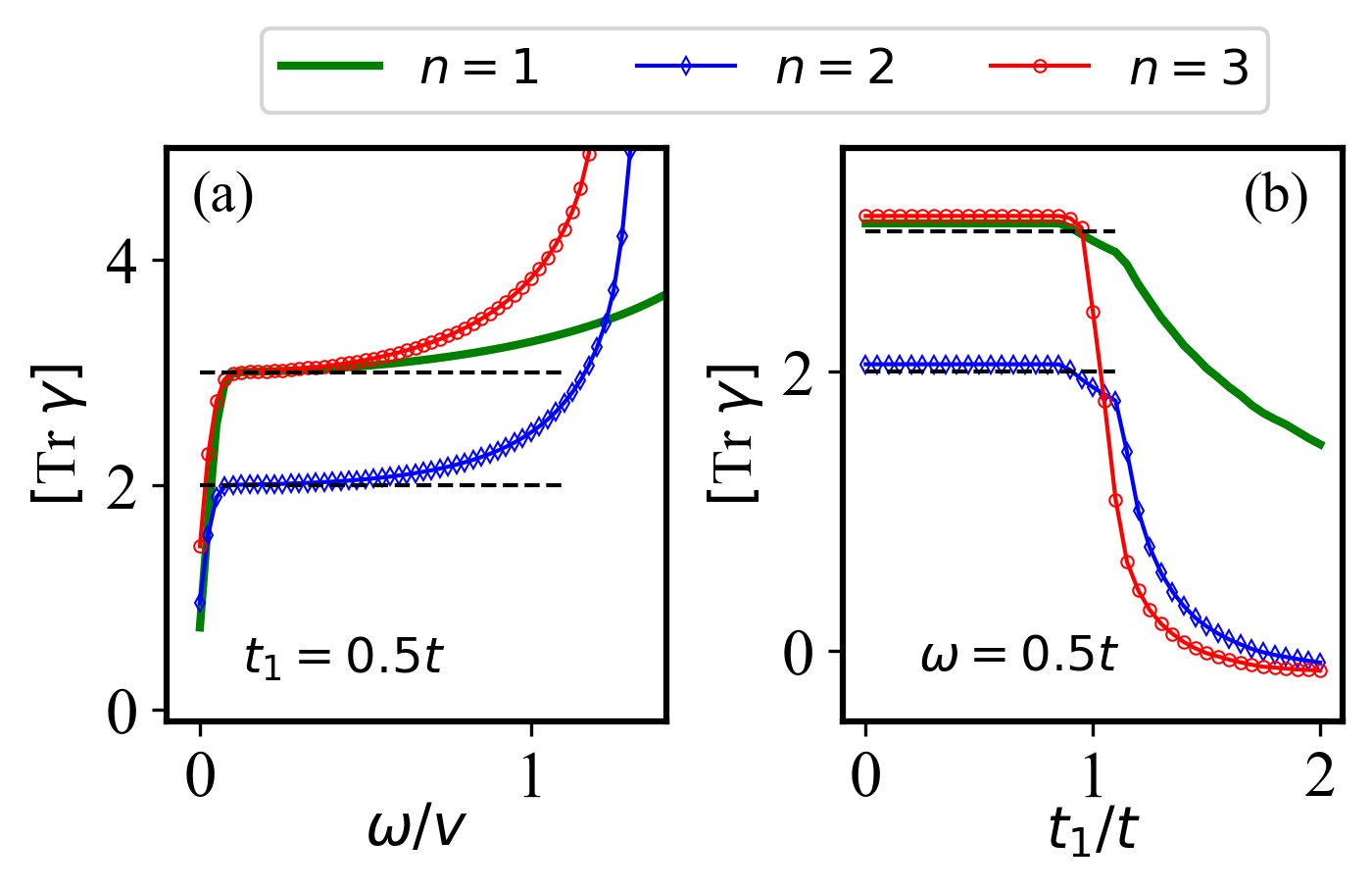}
\caption{Numerical results of $Tr[\gamma]$ based on lattice model of mWSM ($H_L(\mathbf{k})$). Panel (a): frequency-variation of $Tr[\gamma]$ for a fixed tilt strength $t_1=0.5t$. The frequency-independence and quantization signatures can be seen in the lower incident frequency regime while they decrease gradually with increasing light frequency, attributed to the band regulations. Panel (b): variation of $Tr[\gamma]$ with tilt strength ($t_1$) at a fixed frequency $\omega =0.5 t$. It can be seen that the response is tilt-independent in type-I mWSMs and decreases with increasing tilt in type-II mWSMs. Here, we have used $t=t^{\prime}=1$. 
} 
\label{fig2_Lattice}
\end{figure}
\end{center}

Since the model given in Eq.~(\ref{eq_multi1}) lacks a physical ultraviolet energy cutoff, to verify the robustness of the above results, we calculate the OPI response numerically considering a lattice model of mWSM~\cite{Li16}, $H_L(\bm{k}) =N_0 (\bm{k}) \sigma_0 +\bm{N(k)\cdot \sigma}$ with $N_0 (\bm{k})=t_1 \cos k_z \sigma_0$ containing the tilt parameter $t_1$. Here, $N_{z} = t^\prime(2 - \cos{k_x}-\cos{k_y}-\cos{k_z})$, $N_{x}= t \sin{k_x}$ for $n=1$, $t (\cos{k_x}-\cos{k_y})$ for $n=2$, $t\sin{k_x}(3\cos{k_x}-\cos{k_y}-2)$ for $n=3$, and $N_{y}= t \sin{k_y}$ for $n=1$, $t\sin{k_x} \sin{k_y}$ for $n=2$,  $t\sin{k_y}(3\cos{k_x}-\cos{k_y}-2)$ for $n=3$. We depict its variation with light frequency ($\omega$) and tilt parameter ($t_1$) in Fig.~\ref{fig2_Lattice}. It can be seen that the plateaus ($\omega$-independence) quantized to $n$ are destroyed in the high-frequency regime (see Fig.~\ref{fig2_Lattice}(a)). The gradual deviation of ${\rm Tr}\gamma_{\alpha\beta}$ beyond a certain $\omega$ for a type-I mWSM is attributed to the fact that a nonlinearity appears in the band dispersion if one increases $\omega$ beyond a certain cutoff. We note that, like type-I cases, the frequency-independence of ${\rm Tr}\gamma_{\alpha\beta}$ for type-II mWSM lattice models can persist up to a certain frequency~\cite{Cao24}. However, Fig.~\ref{fig2_Lattice}(b) indicates that, unlike a type-I node, the circular OPI is a decreasing function of the tilting parameter in type-II mWSMs. The qualitative differences of the OPI coefficient in type-I and type-II mWSMs can be utilized to distinguish them in experiments.

\begin{center}
\begin{figure}[t]
\includegraphics[width=0.47\textwidth]{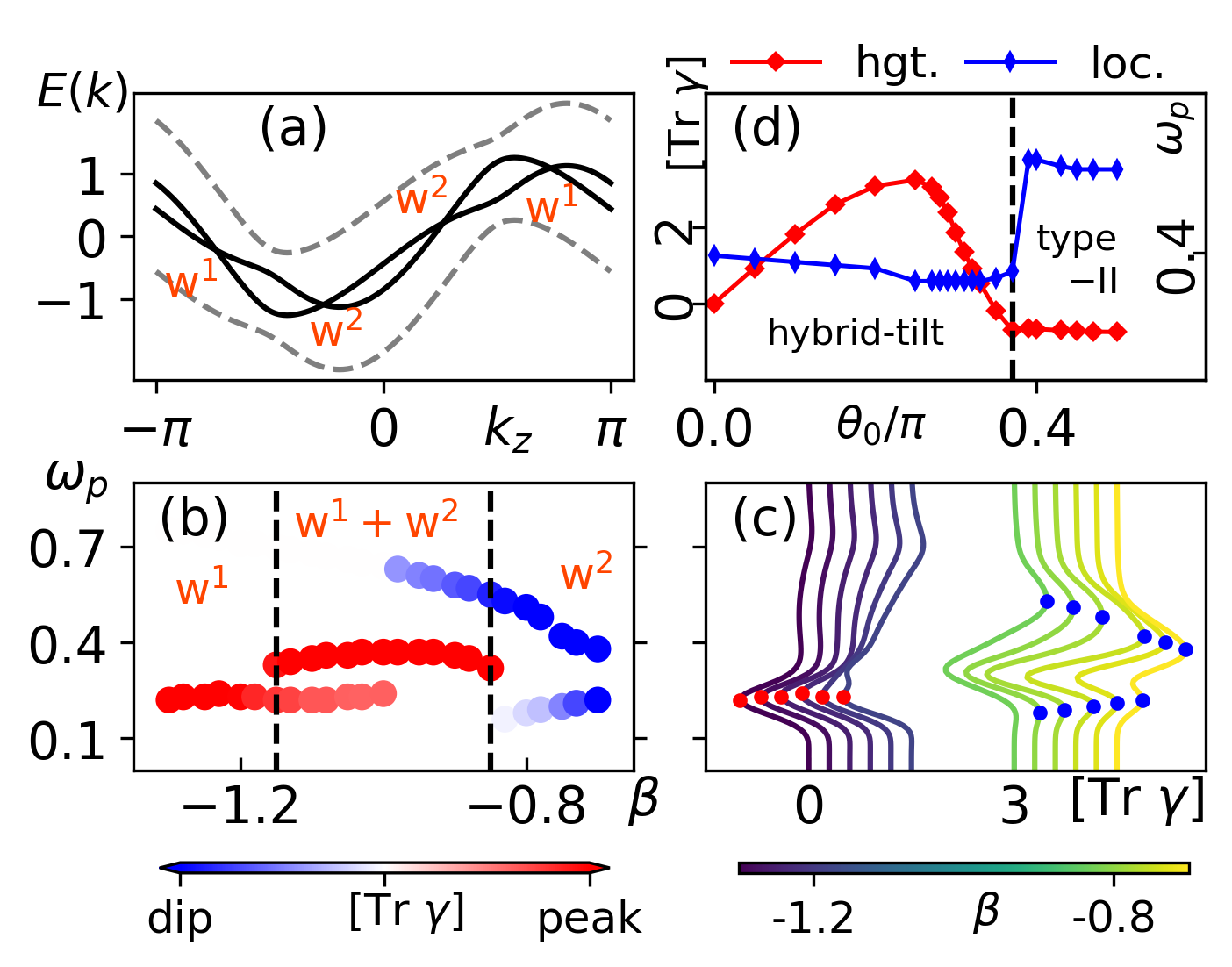}
\caption{Panel (a), band dispersion of the HOWSM based on Hamiltonian ${\rm H}^{\rm HW}_{\mathbf{k}}$~\cite{Hua_2020}. Panel (b) shows the peak (dip) locations in the $\omega-\beta$ parameter space, indicated by color code red (blue) for the response amplitude (height). The dashed lines at $\beta=-1.15, -0.85$ indicate the transition thresholds from the $W^1$ to the $W^1+W^2$ phase and from the $W^1+W^2$ to the $W^2$ phase, respectively. A few line-cut plots associated with the dip/peak locations are shown in Panel (c), where appropriate offsets are applied. Panel (d) plots the value of $Tr\gamma_{\alpha\beta}$ at the peaks, i.e., peak height (peak location), as a function of tilt-tuning parameter $\theta_0$, as given by the red line (blue line). The dashed line $\theta_0 \sim 0.36$ indicates the boundary of the hybrid-tilt and type-II phases. The other parameters are taken from ~\cite{Hua_2020}.} 
\label{fig2_HOWSM}
\end{figure}
\end{center}

{\color{blue}{\em Optical pseudospin injection in HOWSM}}--- To investigate the OPI beyond the first-order WSMs, we consider a time-reversal and inversion broken hybrid-tilted HOWSM, constructed by stacking 2D quadrupole insulators along the $z$-axis and given by the lattice model~\cite{Sayed_2020, Hua_2020}, 
\begin{eqnarray}
&&{\rm H}^{\rm HW}_{\mathbf{k}} = (\beta_x+\cos k_x+\frac{1}{2}\cos k_z)\Gamma_4+\sin k_x \Gamma_3+\sin k_y \Gamma_1
\nonumber \\
&&+(\beta_y+\cos k_y+\frac{1}{2}\cos k_z)\Gamma_2+im_1\Gamma_1 \Gamma_3+m_2 \sin k_z \Gamma_0, \nonumber \\
\label{eq_HOWSM}
\end{eqnarray}
where $\beta_{x(y)}$ represents the intracell coupling along $x(y)$, $\{\Gamma_{\alpha}\}$ are direct products of Pauli matrices $\sigma_i$, $\kappa_i$: $\Gamma_i=-\sigma_2\kappa_i$ ($i=1,2,3$), $\Gamma_0=\sigma_3\kappa_0$ and $\Gamma_4=\sigma_1\kappa_0$. Here, $m_1\, (m_2)$ breaks the time-reversal (inversion) symmetry. 
The resulting band dispersion consisting of four WNs is shown in Fig.~\ref{fig2_HOWSM}(a). Interestingly, varying $m_1$ and the intracell coupling parameter $\beta$ ($\beta_x=\beta_y=\beta$) in ${\rm H}^{\rm HW}_{\mathbf{k}}$, we can achieve regimes containing only $W^1$ (1st-order) nodes, or $W^2$ (2nd-order) nodes, or both the $W^1 +W^2$ (hybrid-order) WNs.

Remarkably, the OPI exhibits a transition of a single-peak feature in the 1st-order Weyl phase to a double-dip feature in the 2nd-order one. It is evident from Fig.~\ref{fig2_HOWSM}(b), where $\omega_p$ indicates peak/dip locations in frequency space. A few line-cuts marked with the corresponding peak/dip locations are given in Fig.~\ref{fig2_HOWSM}(c). Thus, the above peak-to-dip signature in OPI response might help directly probe and discern the 1st-order and 2nd-order WSMs in hybrid-tilted HOWSMs in experiments. 

Furthermore, one can include the tilting of WNs in HOWSMs by adding a term $\propto \sin(k_z-\theta_0)$ to the model in Eq.~(\ref{eq_HOWSM}), where $\theta_0$ is the tilt-tuning parameter~\cite{Sayed_2020}, and thereby providing a chance to realize a complete set of type-I, type-II, and hybrid-tilt phases. Interestingly, we find that both the peak height (dip depth) and the peak (dip) location of ${\rm Tr}\gamma_{\alpha\beta}$ for the HOWSM facilitate the distinction of the hybrid-tilt and type-II HOWSM phases, as shown in Fig.~\ref{fig2_HOWSM}(d). Lastly, we note that the linearized model of HOWSM may be incapable of capturing any distinction in OPI response between 1st-order and 2nd-order WNs as the low-energy behavior is similar for both of them.

{\color{blue}{\em Conclusions}}--- In this work, we investigate circular OPI in multi-WSMs and higher-order WSMs. Remarkably, our work reveals that the OPI in mWSMs can exhibit a frequency-independent quantized (specifically, linearly proportional to $n$) response irrespective of the tilting of WNs, arising from the inherent band linearity in their dispersions. Our argument is supported by the strong directional dependence of OPI response in mWSMs, with only one finite component ($\gamma_{zz}$), unlike ordinary WSMs where all diagonal components remain nonzero. The frequency-independence would have immense importance in the context of magnetic Weyl systems, as CPL can reverse magnetization direction swiftly~\cite{Stanciu07}, offering a rapid data storage method without requiring precise frequency tuning around a resonance. We further unveil that the frequency-independence can be disrupted by transitioning to a higher frequency range owing to the appearance of nonlinearity in the band dispersion beyond a certain $\ omega$ cutoff. Interestingly, unlike a type-I node, the response is a decreasing function of the tilting of a type-II node, thereby enabling the possibility to distinguish them in OPI experiments. 
Investigating different orders of Weyl phases, we show that the OPI response is accompanied by peaks for the first-order Weyl nodes while dips for the second-order Weyl nodes, implying the frequency-dependence even in the low $\omega$ regime. Given the above discussions, the proposed signatures of OPI in mWSMs and HOWSM could lead us to effectively access as well as distinguish different nontrivial Weyl topologies in experiments.

{\color{blue}{\em Acknowledgements}}--- S.G. acknowledges the Ministry of Education, India for research fellowship. C.Z. acknowledges support from the NSF of China (Grants No. 12104043). The work at Los Alamos National Laboratory was carried out under the auspices of the US Department of Energy (DOE) National Nuclear Security Administration under Contract No.89233218CNA000001. It was supported by the LANL LDRD Program, and in part by the Center for Integrated Nanotechnologies, a DOE BES user facility, in partnership with the LANL Institutional Computing Program for computational resources.

\bibliography{OPI}

\end{document}